# A Wideband Multibeam Planar Quasi-Yagi Array Antenna Based on Parasitic Pixel Strips

Jinyang Bi, *Student Member, IEEE*, Fan Qin, *Member, IEEE*, Chao Gu, *Member, IEEE*, and Hailin Zhang, *Member, IEEE*

***Abstract*—This letter presents a low-complexity parasitic pixel strip (PPS) structure for array antenna to generate six radiations with flexible beam control. To realize multibeam operation, three groups of symmetrical strip-shaped pixels are integrated with the feed antenna, the connection state of which can be controlled by PIN diode. Leveraging the reconfigurable parasitic configuration, the coupling path of the antenna can be strategically manipulated, achieving the desired multibeam property. Thus, by utilizing the PPS structure in conjunction with a planar quasi-Yagi antenna, this single antenna can operate at four states with beam direction towards 40°, -40°, 0°, along with a dual-beam at ±45°, within the frequency range from 5.0 to 5.6 GHz. Moreover, a prototype of a 1 × 2 array with the PPS structure is also fabricated to achieve six beams with a scan range of 40°, 20°, 0°, -20°, -40° and a dual-beam at ±30°. Notably, only four PIN diodes are utilized to implement six radiations, verifying that this approach effectively minimizes the excessive use of switches and extends the variety of beams, which provides a low complexity method for multibeam array antenna without complicated beamforming network. These antennas satisfy the benefits of wideband, planner structure, low structural complexity, low cost, and flexible beam control.**

## I. INTRODUCTION

WITH the rapid development of wireless communication technology, the traditional directional antenna can no longer satisfy the requirement of high-quality communication in complex electromagnetic (EM) environments. The potential of pattern reconfigurable antenna allows for switchable beams across extensive coverage, enhancing antenna's performance [1], [2]. By exciting various TM modes, continuous scan beams can be achieved in [3], [4] but lack of design flexibility. Besides, PIN diodes are commonly utilized as radio frequency (RF) switches to implement beam steering, with benefits of flexible design and convenient digital control.

Featuring characteristics like wideband, end-fire radiation, and good robustness, the Yagi or quasi-Yagi antenna has been used to design pattern-reconfigurable antenna [5]. Construct reconfigurable directors and reflectors constitute the majority of methods that achieve beam steering. One category is to reshape the physical length of director and reflector through RF switches [6]-[9]. The other is to create embedded parasitic resonators and change the capacitive or inductive reactance of the fixed-length parasitic elements using switches [10]-[12]. However, only frontward and backward radiated directions can be generally achieved by these antennas. Moreover, based on even-odd mode excitation [13] or length-switching balun [14], the Yagi antenna can generate scan beams. Nevertheless, all the aforementioned designs have difficulty integrating with antenna arrays compactly for multibeam generation, thereby limiting their application range.

Loading parasitic pixels is also an important approach to implement pattern reconfigurability of Yagi antenna [15]-[19]. However, the large number of pixels and switches utilized in pixel antenna exhibit a large structural complexity and large scale. Meanwhile, research on pixel antennas mainly focuses on single-element design, with limited consideration of their integration into antenna arrays. Besides, the beam scanning in Yagi array antenna typically relies on complex beamforming networks such as Rotman lens [20] or Butler matrix [21]-[23]. Hence, realizing a broadly steerable multibeam Yagi array antenna with low complexity remains a challenge.

This letter presents a reconfigurable pixel array antenna by integrating the diode-controlled parasitic pixel strips (PPSs) and 1 × 2 quasi-Yagi array together. Only four PIN diodes are utilized to achieve the low-complexity multibeam design for array antenna without complex feeding networks. Based on the parasitic structure, the single antenna and 1 × 2 array have been fabricated to realize the beam-scanning and dual-beam radiations, validating an effective low-complexity multibeam design and broad application on wireless communication.

## II. SINGLE ANTENNA DESIGN

### *A. Antenna Structure*

The schematic of the proposed single antenna is illustrated in Fig. 1, which consists of two metal layers and one dielectric substrate. The top side of the substrate comprises a microstrip feed line, a half-wavelength dipole driver with a director, three pairs of symmetrical strip-shaped parasitic pixels connected by four PIN diodes, and dedicated biasing lines. The bottom side is a truncated ground, serving as a reflector. The PPSs are placed respectively in front of, to the left, and to the right of the feed. Four PIN diodes are respectively soldered in the gaps between every two adjacent strips. Rogers 4003C is adopted as the dielectric substrate with a relative permittivity of 3.55, loss tangent of 0.0027, and thickness of 0.508 mm.

### *B. Principle of Pattern Diversity*

Fig. 2 illustrates the electric field distributions of the single antenna at 5.3 GHz under various operating modes. For the quasi-Yagi antenna with a single director and without PPS structures, the electric field is mainly coupled from the driven dipole to the director and primarily concentrates at the left and right edges of the strip, directing the energy to radiate at the

Fan Qin, Jinyang Bi, Hailin Zhang are with the School of Telecommunications Engineering, Xidian University, Xi'an 710071, China (e-mail: fqin@xidian.edu.cn; jybi@stu.xidian.edu.cn; hlzhang@xidian.edu.cn).

Chao Gu is with the ECIT Institute, Queen's University Belfast, BT3 9DT Belfast, U.K. (e-mail: chao.gu@qub.ac.uk).

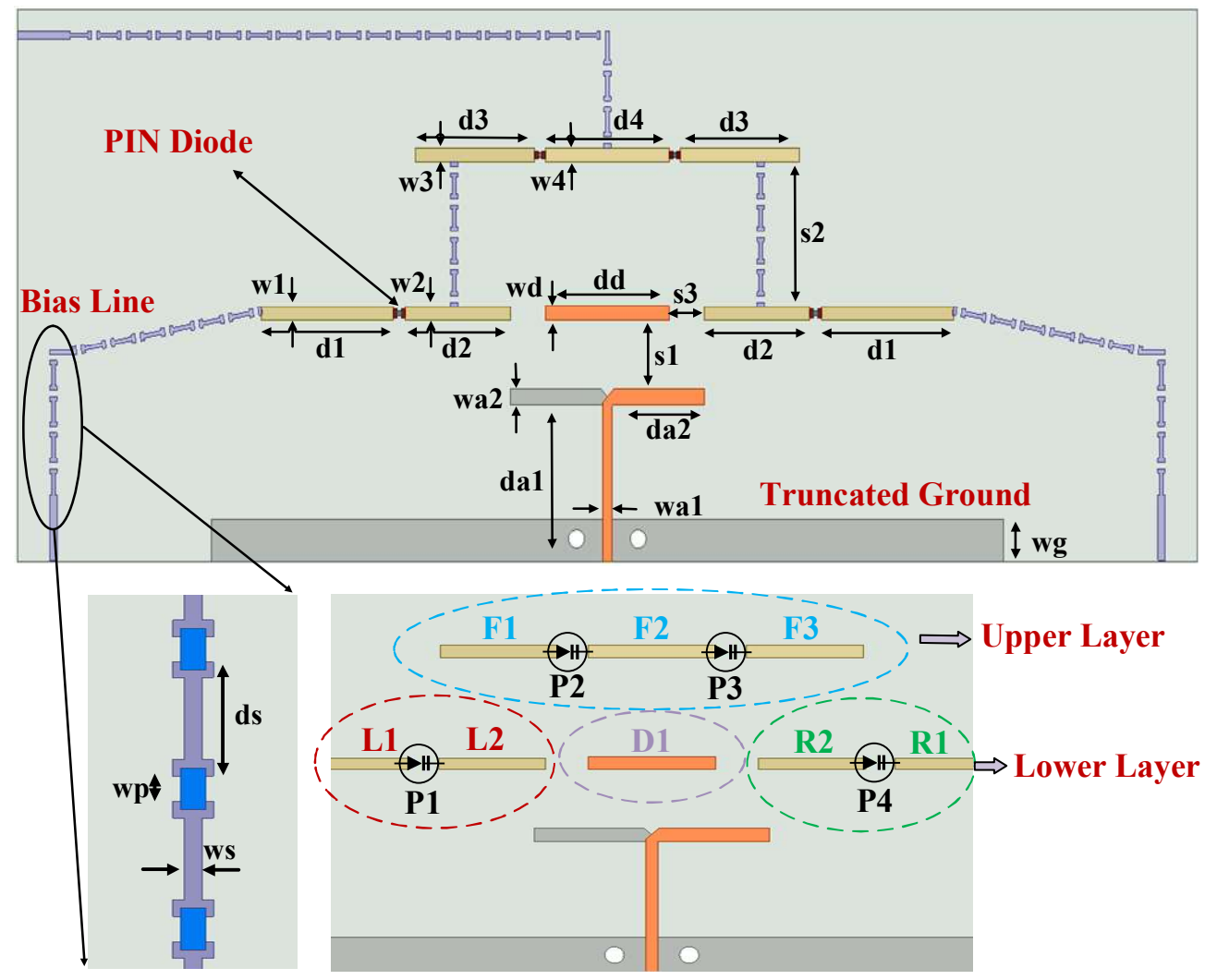

Fig. 1. Topology of the single antenna. Dimensions are as follows: d1 = 15, w1 = 1.3, d2 = 12, w2 = 1.3, d3 = 13.5, w3 = 1.4, d4 = 14.1 w4 = 1.4, dd = 14.1, wd = 1.5, da1 = 14.7, wa1 = 1.1, da2 = 10.5, wa2 = 1.5, wg = 4, ds = 2.6, ws = 0.35, wl = 0.6, s1 = 6.5, s2 = 13.6, s3 = 4. (Units: mm).

end-fire direction, as depicted in Fig. 2(a).

In State I, with all PIN diodes off and all gaps open, the EM energy is coupled from the dipole to the edges of the director D1 and all PPS structures. The separate strip F2 performs as the second layer director, and the energy is mainly directed to the end-fire direction, as depicted in Fig. 2(b).

In State II, diodes P1 and P2 are turned on while P3 and P4 turned off. As shown in Fig. 2(c), except director D1, the EM energy is primarily coupled to the edges of separate strips F3, R1, and R2 in the upper-right direction. Since strips L1-L2 and F1-F2 are connected without gaps, minimal coupling occurs in these strips. This shift in coupling path and primary coupling points results in a deflection of the radiation pattern. Only a small amount of energy couples at leftmost end of L1, which manifests as a sidelobe in the far-field radiation pattern. From another perspective, connected strips L1-L2 and F1-F2 act as reflectors, obstructing end-fire radiation and redirecting the beam towards the upper-right direction. Thus, a right-titled beam directed at 40° can be obtained.

In State III, diodes P3 and P4 are activated while P1 and P2 turned off. The coupling path operates oppositely to the State II, redirecting a -40° left-titled beam, as depicted in Fig. 2(d).

In State IV, with all diodes turned on and all gaps closed, the radiated energy is weakly coupled to the connected strips L1-L2, R1-R2, and F1-F2-F3. The F1-F2-F3, unable to couple energy, blocks its propagation path and performs as a reflector from another perspective. Thus, this energy accumulates and oscillates between the PPS structures and the ground before eventually radiating through the gaps of parasitic components, forming a dual-beam of ±45°, as shown in Fig. 2(e).

The arrangements and electrical sizes of parasitic strips are precisely optimized to achieve parasitic coupling manipulation. By altering the coupling path and coupling point positions, the coupled energy is directed to desired directions. The simulated radiation patterns are listed in Fig. 3. In total, only four diodes are utilized to realize four beam modes, ensuring a minimum quantity of diodes used, low cost, and design simplification.

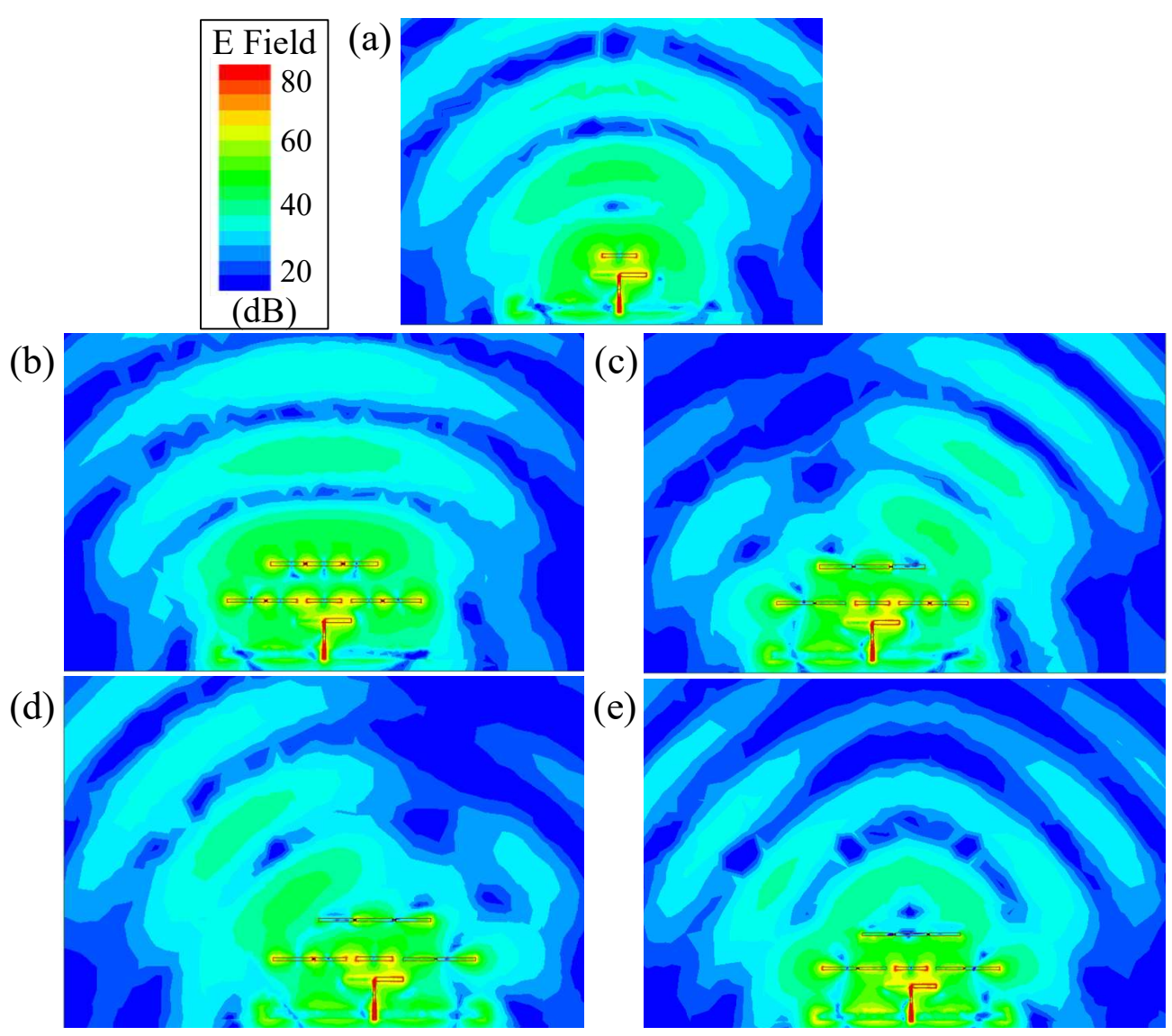

Fig. 2. The E-field distributions of the single antenna (a) without PPSs (b) in state I: end-fire beam (b) in state II: right-titled beam of 40° (c) in state III: left-titled beam of -40° (d) in state IV: dual-beam.

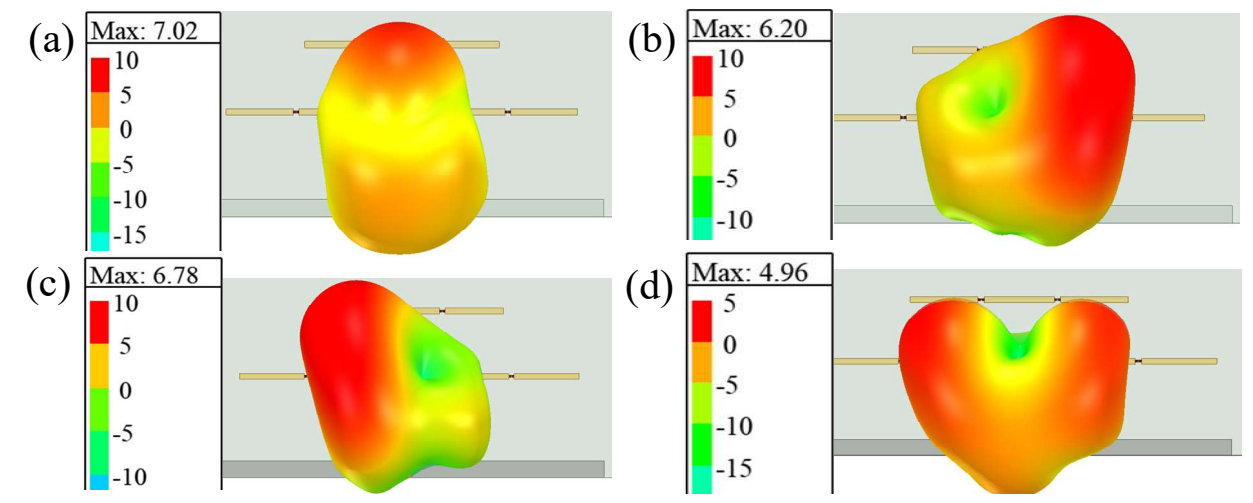

Fig. 3. Simulated radiation patterns of the single antenna at 5.3 GHz (a) I. end-fire beam (b) II. right-titled beam (c) III. left-titled beam (d) IV. dual-beam.

### C. DC Biasing Network

The bias lines, located on the top side of the substrate, use very thin (0.3 mm width) metal wires and split them into short non-resonant sections by means of RF chokes [16]-[19]. Due to the bias lines arranged in the propagation path of the EM wave radiated by the feed, this design minimizes the radiation energy coupling on the bias lines and prevents them from converting to reflectors, ensuring almost no effect on radiation performance. MACOM MADP-000907-14020 PIN diode was chosen as the suitable component [24]. For simulation of the ON state, it can be approximated as a resistance of 7.8 Ω and an inductance of 30 pH in series. In the OFF state, it exhibits a capacitance of 0.025 pF and an inductance of 30 pH in series. A 20 nH surface inductor is selected as the RF choke.

## III. 1 × 2 Array Design

Based on the above PPS structure, a 1 × 2 quasi-Yagi array antenna is also constructed with an overall dimension of 150 mm × 60 mm, as shown in Fig. 4. This array consists of two feed elements with two directors, four pairs of parasitic pixels, four PIN diodes, and biasing lines. Besides four working states similar to the single antenna, the number of scan beams can be further increased by either simultaneous or separate excitation of the array's two feed ports. Table I lists the working modes with corresponding diode and feed port states.

Fig. 5 depicts the E-field distributions of the array under six

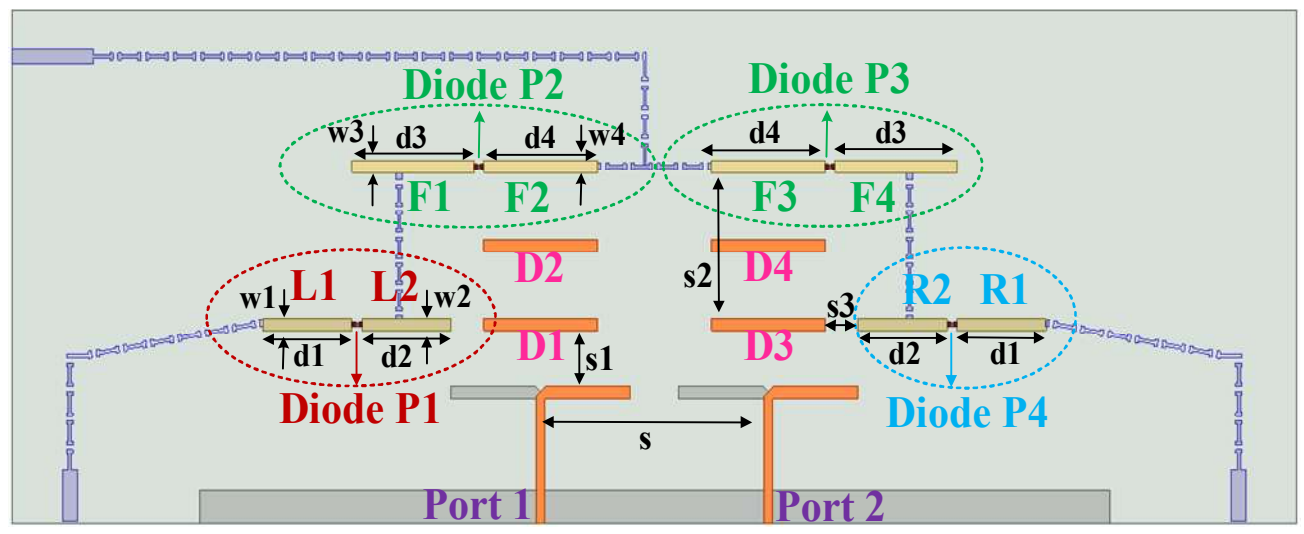

Fig. 4. The geometry of the 1 × 2 array antenna. Dimensions are as follows: d1 = 11.5, w1 = 1.3, d2 = 11, w2 = 1.3, d3 = 15, w3 = 1.4, d4 = 14, w4 = 1.4, s = 28.3, s1 = 6.5, s2 = 17.1, s3 = 4 (Units: mm).

TABLE I
SIX BEAM MODES CORRESPONDING TO VARIOUS STATES OF PIN DIODES

| Six Beam States | Diodes Modes (1-4) | Excited Feeding Ports |
|---|---|---|
| I. End-fire Beam | 0 0 0 0 | 1, 2 |
| II. Right-titled Beam of 20° | 1 1 0 0 | 1, 2 |
| III. Left-titled Beam of -20° | 0 0 1 1 | 1, 2 |
| IV. Dual-Beam | 1 1 1 1 | 1, 2 |
| V. Right-titled Beam of 40° | 1 1 0 0 | 1 |
| VI. Left-titled Beam of -40° | 0 0 1 1 | 2 |

*'0' represents reverse-biased mode, and '1' represents forward-biased mode.

modes. In Figs 5(a) and (d), the working principles of states I and IV closely resemble those of the end-fire and dual-beam modes of the single antenna, respectively, with key distinction being that the feed is upgraded from a dipole to a 1 × 2 array.

For States II and III of the left- and right-titled beams of 20°, both two feed ports operate simultaneously. In State II, due to the diodes P1 and P2 turned on, the energy radiated by port 1 is weakly coupled to the connected strips L1-L2 and F1-F2, preventing end-fire radiation and instead radiating through the upper-right gap, as shown in Fig. 5(b). Meanwhile, the energy radiated by port 2 is typically coupled to its end-fire direction. These two beams merge to form a right-titled beam deviating from the end-fire direction by 20°. From another perspective, the connected strips L1-L2 and F1-F2 function as reflectors. State III operates in direct contrast to State II.

For States V and VI of the left- and right-titled beams of 40°, only one feed port is excited. In State V, as shown in Fig. 5(e), with PIN diodes P1 and P2 turned on, the energy radiated by port 1 is weakly coupled to the connected L1-L2 and F1-F2, but instead primarily radiates from the upper-right gap. Unlike state II, where the end-fire beam from port 2 has an impact, in this case, there is no such influence since port 2 is not excited, and the right-tilted beam is oriented at 40°. State VI operates in direct contrast to State V.

Based on the above analysis from the perspective of mutual coupling, the antenna's coupling path is effectively altered by switching diodes on or off, enabling a manipulation of energy to reradiate in the desired direction. To verify this design, Fig. 6 depicts the radiation patterns for six beam modes at 5.3 GHz.

## IV. SIMULATED AND MEASURED RESULTS

Shown in Fig. 7, two prototypes of the single antenna and array were fabricated and measured in an anechoic chamber. The measured reflection coefficients of the single antenna are plotted in Fig. 8(a). Each beam has a bandwidth covering the frequency range from 5.0 to 5.9 GHz, with a 17.0% relative bandwidth. The isolation of the array is also investigated in

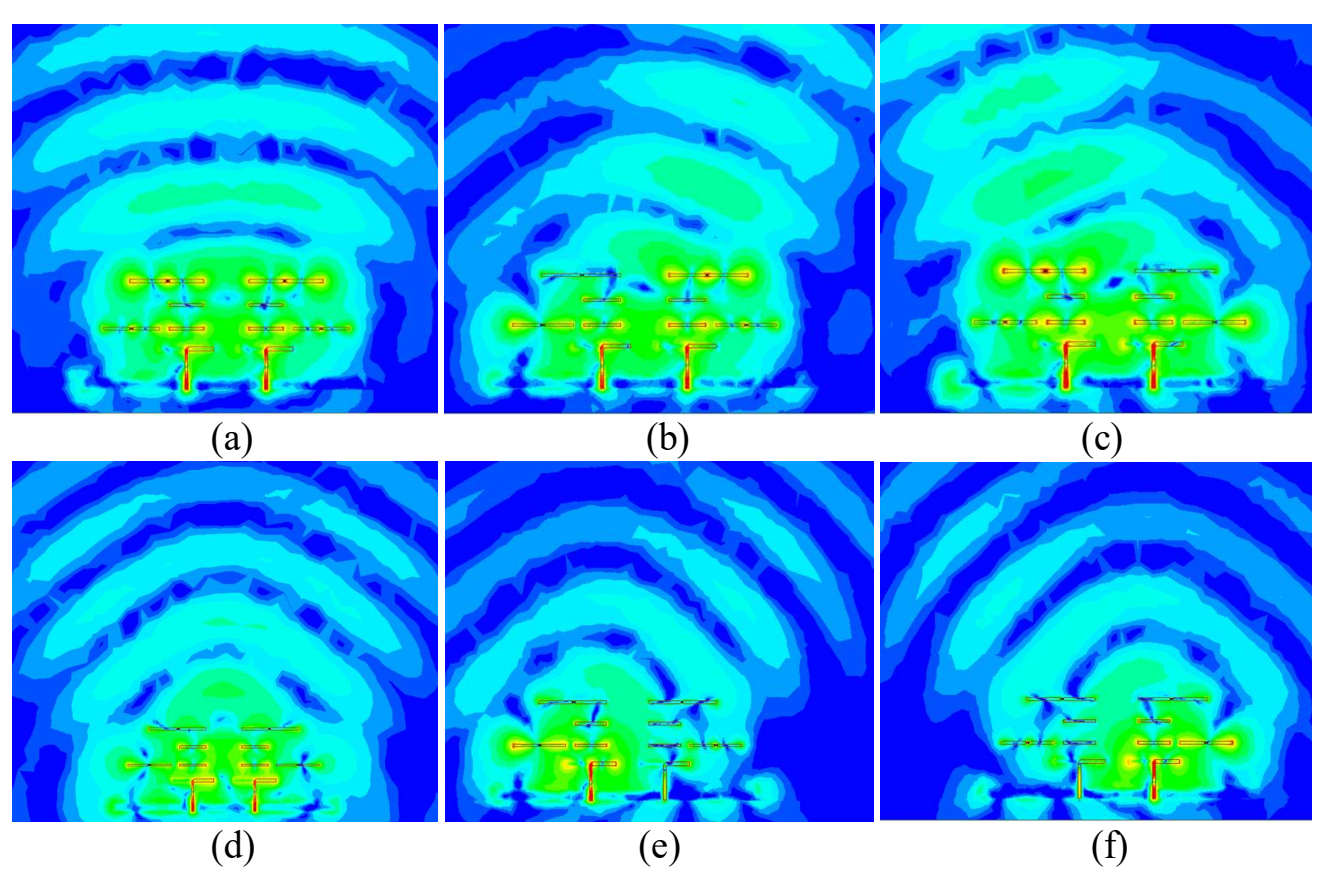

Fig. 5. E-field distributions at 5.3 GHz for six radiation modes (a) I. end-fire beam (b) II. right-titled beam of 20° (c) III. left-titled beam of -20°(d) IV. dual beam of ±30° (e) V. right-titled beam of 40° (f) VI. left-titled beam of -40°.

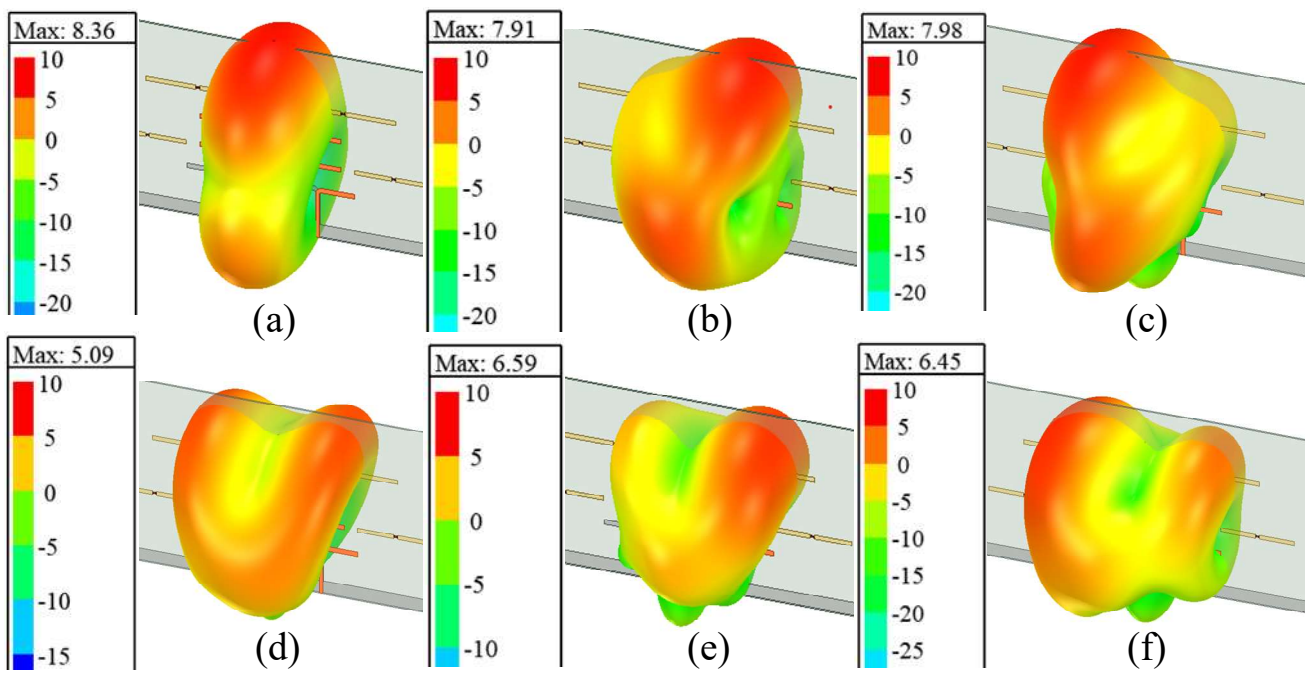

Fig. 6. Simulated 3-D patterns at 5.3 GHz corresponding to six beam modes (a) end-fire beam (b) right-titled beam of 20° (c) left-titled beam of -20°(d) dual beam (e) right-titled beam of 40° (f) left-titled beam of -40°.

Fig. 8(b). The isolation between the two ports is better than 15 dB across the whole bandwidth and 23 dB at 5.3GHz.

The measured normalized E-plane radiation patterns of the single antenna at 5.0, 5.3, and 5.6 GHz are plotted in Fig. 9. It achieves a scan range of -2°, 39°, and -40°, and dual-beam at ±45°, maintaining a wideband range from 5.0 to 5.6 GHz. The measured gains of four states are 5.17 - 7.09 dBi, 4.74 - 6.01 dBi, 5.52 - 6.37 dBi, and 2.68 - 4.47 dBi, respectively, across from 5.0 to 5.6 GHz band. The gain variations are stable, less than 2.35 dB, across the whole scan range over the entire bandwidth. The simulated peak gains are 7.33, 6.47, 6.90, and 5.08 dBi, respectively. The additional loss between measured and simulated results is mainly attributed to substrate losses, test errors, and assembly errors. Since the diodes only impact the EM wave propagation path rather than directly interacting with the feeding network or antenna aperture, their ohmic loss remains minimal. The measured cross-pol levels are less than -18 dB inside the main lobe. The measured front-to-back ratio (FBR) at 5.3 GHz in end-fire mode is 9.9 dB.

The measured normalized E- and H-plane radiation patterns of the array are plotted in Figs. 10 and 11(a), achieving beam scanning with scan angles of 0°, 20°, -20°, 38°, and -41°, as well as a dual-beam at 30° and -31° at 5.3 GHz. The measured gains for six beams at 5.3 GHz are 8.11, 7.65, 7.58, 6.30, 6.23, and 4.69 dBi, respectively. Additionally, the measured cross-pol levels are less than -20 dB inside the main lobe, while the measured FBR at 5.3 GHz in end-fire mode is 14.0 dB.

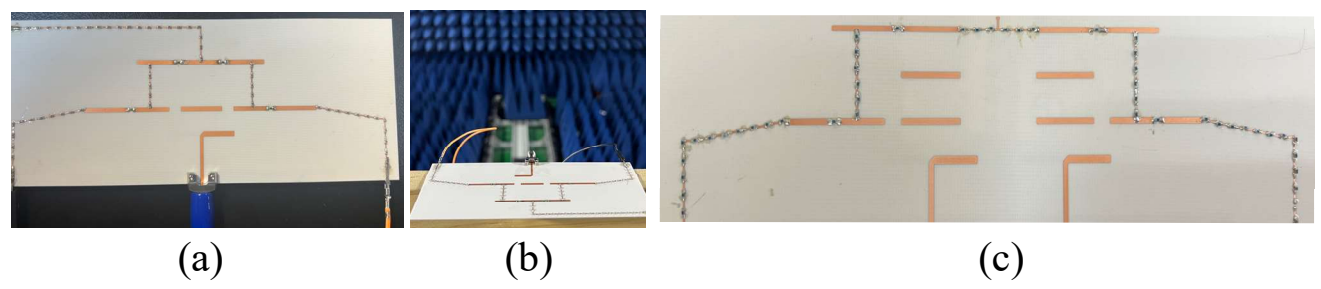

(a) (b) (c)

Fig. 7. Photographs of the prototypes: (a) top view of the single antenna (b) single antenna measurement in an anechoic chamber (c) top view of the array.

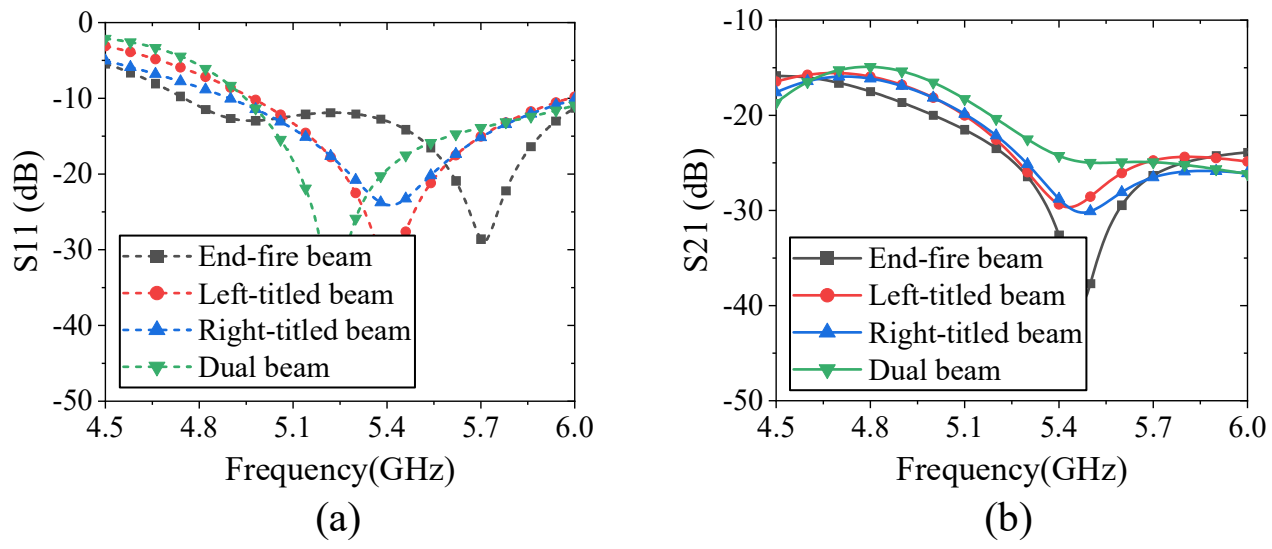

Fig. 8. (b) $S_{11}$ of the single antenna. (a) Isolation characteristic of the array.

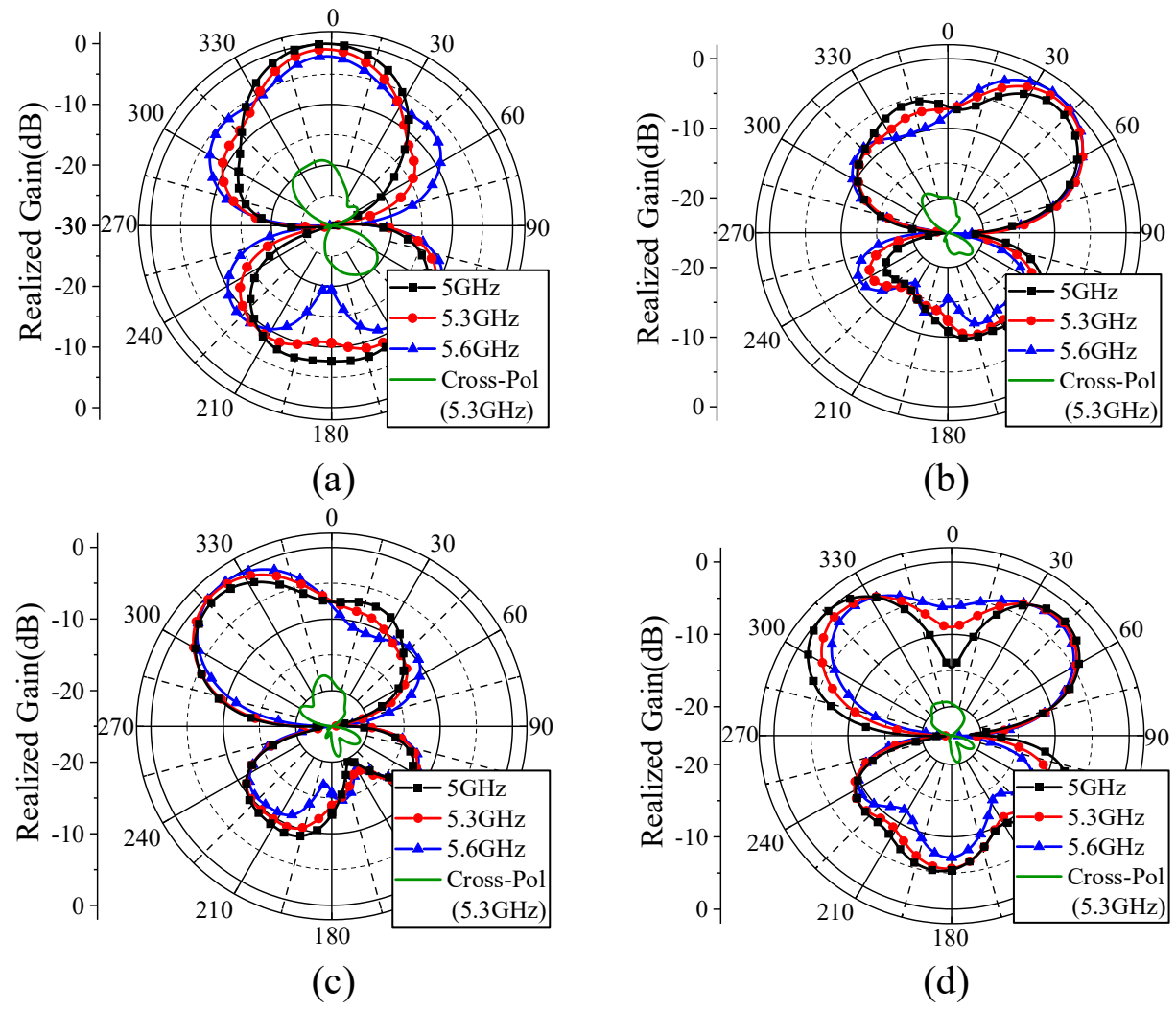

Fig. 9. Measured E-plane radiation patterns of the single antenna. (a) end-fire beam (b) right-titled beam (c) left-titled beam (d) dual-beam.

To illustrate the relationship between the dual-layer PPSs and beam orientation, Fig. 11(b) compares radiation patterns between the individual action of lower PPS and the combined action of dual-layer PPSs. The beam scanning range of the 1 × 2 array narrows from 40° to 30° with only the lower layer active. All PPS parameters have been optimized concurrently to balance the peak scan coverage and high directivity.

To highlight the novelty of our array, a comparison with the existing multibeam antennas is listed in Table II. Unlike prior research, our design presents a low-complexity approach for array antenna to achieve multibeam property, which facilitates flexible beam switching across six radiation states, avoiding extensive use of PIN diodes and realizing a high utilization efficiency. On the one hand, unlike traditional multi-beam array antennas that rely on complex beamforming networks, this design simplifies the structural complexity significantly and offers advantages of low cost, compact size, and flexible digital control. On the other hand, it outperforms multibeam single antennas by providing higher gain, fewer switches, and enhanced beam pointing capabilities. Meanwhile, it features a low profile and good scan bandwidth, highlighting potential for broader applications in wireless communication systems.

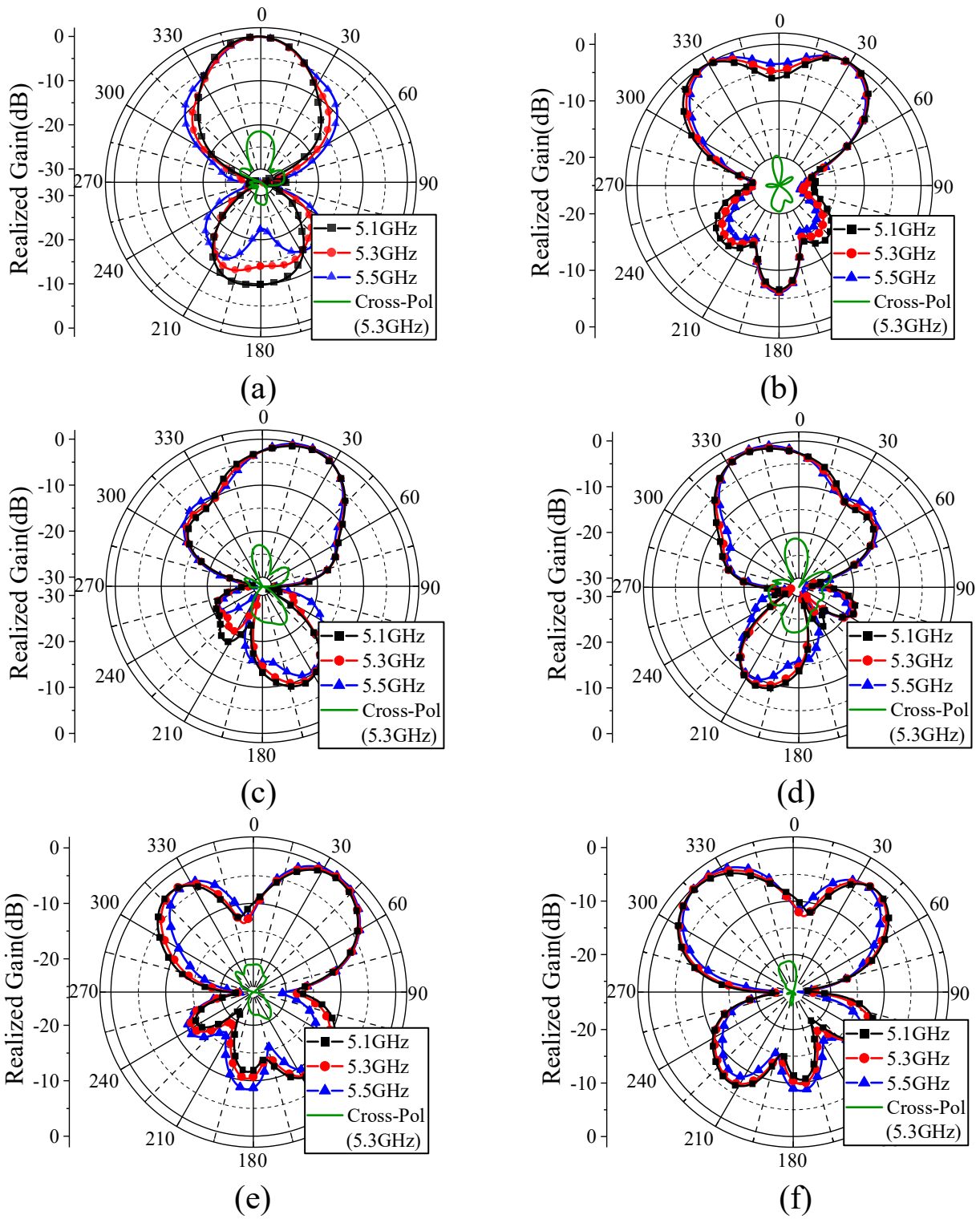

Fig. 10. Measured E-plane radiation patterns of the 1 × 2 antenna array at (a) 0° (b) ±30° (c) 20° (d) -20° (e) 38° (f) -41°.

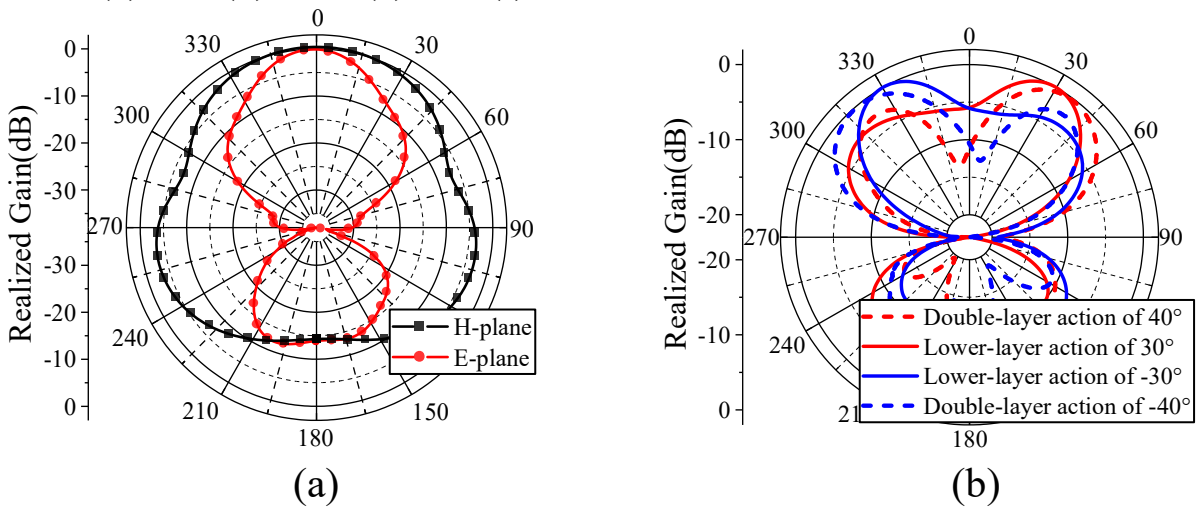

Fig. 11(a). H-plane radiation pattern of the 1 × 2 array at 5.3 GHz. (b) E-plane radiation patterns under various PPS layer actions of the 1 × 2 array.

TABLE II
COMPARISON OF THE PROPOSED ARRAY AND RELATED WORKS

| Ref | Gain (dB) | Scan BW | Size ($\lambda_0^2$) | No. of Mode | No. of Diode | No. of unit | Profile ($\lambda_0$) |
|---|---|---|---|---|---|---|---|
| [8] | 6.4 | 14.8 | 0.7×0.54 | 3 | 16 | 1 | 0.68 |
| [9] | 7.3 | 5.9 | 1.2×1.2 | 4 | 8 | 1 | 0.08 |
| [14] | 10.0 | 3.8 | 1.9×1.2 | 3 | 24 | 1 | 0.02 |
| [15] | 8.0 | / | 4.9× 2.8 | 2 | 2 | 1 | 0.05 |
| [19] | 7.5 | / | 2.5×1.0 | 3 | 22 | 1 | 0.32 |
| [20] | 9.4 | / | 14.4×10.3 | 7 | / | 8 | 0.01 |
| [21] | 5.6 | 14.3 | / | 7 | / | 4 | 0.07 |
| Prop. | 8.1 | 7.5 | 2.6×1.1 | 6 | 4 | 2 | 0.01 |

## V. CONCLUSION

In this letter, we propose a low-complexity multibeam array antenna with six switchable beams using only four PIN diodes. The beam switching capability is evoked by manipulating the coupling path using PIN diodes and reradiating the coupled energy to desired multibeam directions. Based on the principle, a prototype of the 1 × 2 array has been optimized to reach a pattern diversity of 0°, +/-20°, +/-40°, and dual-beam of ±30°. This design offers the benefits of a planar structure, wideband operation, simple design, and flexible control, which enhances multibeam capability of array antenna with fewer PIN diodes.

## REFERENCES

[1] W. Lin, H. Wong and R. W. Ziolkowski, "Wideband Pattern-Reconfigurable Antenna With Switchable Broadside and Conical Beams," in *IEEE Antennas and Wireless Propagation Letters*, vol. 16, pp. 2638-2641, 2017.

[2] S. -L. Chen, P. -Y. Qin, W. Lin and Y. J. Guo, "Pattern-Reconfigurable Antenna With Five Switchable Beams in Elevation Plane," in *IEEE Antennas and Wireless Propagation Letters*, vol. 17, no. 3, pp. 454-457, March 2018.

[3] F. T. Çelik, L. Alatan and O. A. Civi, "A Pattern Reconfigurable Multi-Mode Antenna Based on Circular Disk and Ring-Shaped Resonators," 2022 IEEE International Symposium on Antennas and Propagation and USNC-URSI Radio Science Meeting (AP-S/URSI), Denver, CO, USA, 2022, pp. 405-406.

[4] T. Q. Tran and S. K. Sharma, "Radiation Characteristics of a Multimode Concentric Circular Microstrip Patch Antenna by Controlling Amplitude and Phase of Modes," in *IEEE Transactions on Antennas and Propagation*, vol. 60, no. 3, pp. 1601-1605, March 2012.

[5] C. Deng, W. Yu and K. Sarabandi, "A Compact Vertically Polarized Fully Metallic Quasi-Yagi Antenna With High Endfire Gain," in *IEEE Transactions on Antennas and Propagation*, vol. 70, no. 7, pp. 5959-5964, July 2022.

[6] X. Ding and B. -Z. Wang, "A Novel Wideband Antenna With Reconfigurable Broadside and Endfire Patterns," in *IEEE Antennas and Wireless Propagation Letters*, vol. 12, pp. 995-998, 2013.

[7] S. Zhang, G. H. Huff, J. Feng and J. T. Bernhard, "A pattern reconfigurable microstrip parasitic array," in *IEEE Transactions on Antennas and Propagation*, vol. 52, no. 10, pp. 2773-2776, Oct. 2004.

[8] J. Ren, X. Yang, J. Yin and Y. Yin, "A Novel Antenna with Reconfigurable Patterns Using H-Shaped Structures," in *IEEE Antennas and Wireless Propagation Letters*, vol. 14, pp. 915-918, Dec. 2015.

[9] W. -Q. Deng, X. -S. Yang, C. -S. Shen, J. Zhao and B. -Z. Wang, "A Dual-Polarized Pattern Reconfigurable Yagi Patch Antenna for Microbase Stations," in *IEEE Transactions on Antennas and Propagation*, vol. 65, no. 10, pp. 5095-5102, Oct. 2017.

[10] J. Lu, H. C. Zhang, P. H. He, M. Wang and T. J. Cui, "Pattern Reconfigurable Yagi Antenna Based on Active Corrugated Stripline," in *IEEE Transactions on Antennas and Propagation*, vol. 71, no. 1, pp. 1011-1016, Jan. 2023.

[11] X. -S. Yang, B. -Z. Wang, W. Wu and S. Xiao, "Yagi Patch Antenna With Dual-Band and Pattern Reconfigurable Characteristics," in *IEEE Antennas and Wireless Propagation Letters*, vol. 6, pp. 168-171, 2007.

[12] F. Farzami, S. Khaledian, B. Smida and D. Erricolo, "Pattern-Reconfigurable Printed Dipole Antenna Using Loaded Parasitic Elements," in *IEEE Antennas and Wireless Propagation Letters*, vol. 16, pp. 1151-1154, 2017.

[13] Z. Wang, Y. Ning and Y. Dong, "Compact Shared Aperture Quasi-Yagi Antenna With Pattern Diversity for 5G-NR Applications," in *IEEE Transactions on Antennas and Propagation*, vol. 69, no. 7, pp. 4178-4183, July 2021.

[14] P. -Y. Qin, Y. J. Guo and C. Ding, "A Beam Switching Quasi-Yagi Dipole Antenna," in *IEEE Transactions on Antennas and Propagation*, vol. 61, no. 10, pp. 4891-4899, Oct. 2013.

[15] S. Tang, Y. Zhang, Z. Han, C. -Y. Chiu and R. Murch, "A Pattern-Reconfigurable Antenna for Single-RF 5G Millimeter-Wave Communications," in *IEEE Antennas and Wireless Propagation Letters*, vol. 20, no. 12, pp. 2344-2348, Dec. 2021.

[16] P. Lotfi, S. Soltani and R. D. Murch, "Printed Endfire Beam-Steerable Pixel Antenna," in *IEEE Transactions on Antennas and Propagation*, vol. 65, no. 8, pp. 3913-3923, Aug. 2017.

[17] Z. Li, E. Ahmed, A. M. Eltawil and B. A. Cetiner, "A Beam-Steering Reconfigurable Antenna for WLAN Applications," in IEEE Transactions on Antennas and Propagation, vol. 63, no. 1, pp. 24-32, Jan. 2015.

[18] X. Yuan et al., "A Parasitic Layer-Based Reconfigurable Antenna Design by Multi-Objective Optimization," in IEEE Transactions on Antennas and Propagation, vol. 60, no. 6, pp. 2690-2701, June 2012.

[19] D. Subramaniam *et al*., "High Gain Beam-Steerable Reconfigurable Antenna using Combined Pixel and Parasitic Arrays," *2020 50th European Microwave Conference (EuMC)*, Utrecht, Netherlands, 2021, pp. 718-721.

[20] M. Heino, C. Icheln, J. Haarla and K. Haneda, "PCB-Based Design of a Beamsteerable Array With High-Gain Antennas and a Rotman Lens at 28 GHz," in *IEEE Antennas and Wireless Propagation Letters*, vol. 19, no. 10, pp. 1754-1758, Oct. 2020.

[21] D. Wang, E. Polat, H. Tesmer, H. Maune and R. Jakoby, "Switched and Steered Beam End-Fire Antenna Array Fed by Wideband Via-Less Butler Matrix and Tunable Phase Shifters Based on Liquid Crystal Technology," in *IEEE Transactions on Antennas and Propagation*, vol. 70, no. 7, pp. 5383-5392, July 2022.

[22] H. -J. Dong, Y. -B. Kim and H. L. Lee, "Reconfigurable Quad-Polarization Switched Beamforming Antenna With Crossed Inverted-V Array and Dual-Butler Matrix," in *IEEE Transactions on Antennas and Propagation*, vol. 70, no. 4, pp. 2708-2716, April 2022.

[23] Z. Hu, W. Wang, Z. Shen and W. Wu, "Low-Profile Helical Quasi-Yagi Antenna Array With Multibeams at the Endfire Direction," in *IEEE Antennas and Wireless Propagation Letters*, vol. 16, pp. 1241-1244, 2017.

[24] MACOM. MADP-000907-14020 SPICE Model. Accessed: May. 5, 2023. [Online]. Available: MADP-000907-14020x.pdf (macom.com).